# A BICONTINUOUS STRUCTURE IN SOME SYSTEMS WITH CUBIC MESOPHASES

A.N. Yakunin

Karpov Institute of Physical Chemistry, 103064 Moscow, Russia

*Devoted to memory of*
*Alexander T. Dembo*

**Abstract.** A cubic structure of polymer colloid complexes is studied. The technique of the research includes i) an analysis of well-known literature SAXS data; on this base, at some assumptions, ii) constructing a simple model to estimate geometric structure parameters and to receive a simulated scattering curve, iii) calculations of energetic parameters by help of the analytical fluctuation theory of the phase transitions; iv) comparing the model with the real structure obtained from the SAXS data by the electron density distribution reconstruction method. In the presence of an effective field the 2-nd order phase transition transforms to the 1-st order one in the vicinity of the Lifshitz point. As a result, a bicontinuous structure in cubic mesophases is formed.

**INTRODUCTION**

Supramolecular chemistry and chemical engineering are regarded as the main directions of the modern material science [1]. Self-assembly and self-organizing processes of supramolecular nanostructures are extensively used at the creation of new materials sensitive to various external influences such as changes of temperature, pressure, electrical or magnetic field, chemical nature of an environment, etc [2-6]. The molecular recognition of (bio-) polymer fragments, their ordering and the self-assembly of composing elements result in a spontaneous formation of functional supramolecular structures owing to weak non covalent interactions as van der Waals and electrostatic forces, hydrogen bonding, hydrophobic interactions. Frequently one can observe that the variety of the forms of supramolecular objects is mainly defined by shapes of elementary units [7].

However, the form of a self-organizing supramolecular structure and lattice types depend on not only the chemical nature but external conditions. For example, in well enough investigated lyotropic systems (aqueous surfactant solutions), despite of a complex character of the phase diagrams, temperature of the solution vs. the surfactant concentration [8, 9], it is possible to note the following basic transitions with the surfactant concentration growth

$$\text{I - CP - HEX - LAM} \qquad (1)$$

where I is an isotropic liquid, CP is a cubic phase, as an example, of the $Pm\bar{3}n$ symmetry [8], HEX is a hexagonal phase, LAM is a lamellar phase. CP of the $Ia\bar{3}d$ symmetry can exist in some temperature and concentration range between HEX and LAM [8]; under certain conditions bicontinuous structures of cubic symmetry, limited by (saddle) surfaces of the zero average curvature and negative Gauss one, may become steady [8, 10].

The previous experimental researches of polymer colloid complexes (PCC), consisting of a linear or network polyelectrolyte, an oppositely charged surfactant and water, have resulted in the following observations: various three-dimensional (3D) mesophases such as face centered cubic (FCC) [11], primitive cubic [12], cubic [13] can be obtained in a polyelectrolyte sample, which has no regular structure originally, after the interaction with the surfactants. The change of the water content causes the reversible transition HEX ↔ LAM [14], observable directly during experimental studies of one sample, whereas the transition CP in HEX due to the formation of the complex between a cation gel and sodium alkyl sulfates [13] takes place in different samples with increasing the hydrophobic surfactant radical length measured in the number of $CH_2$ groups from 10 to 12 and more.

Various one-, two-dimensional (2D) and 3D mesophases have been found and investigated by structural methods in thermotropic and lyotropic materials: biopolymers [8], dendrimers [15, 16], block copolymers and their blends [17-22], PCC of linear and network polyelectrolytes with oppositely charged surfactants, the pioneer work [23] fulfilled by A.T. Dembo and his colleagues has stimulated the subsequent extensive researches [11-14, 24-32]. As a result of all these studies, new nanostructures which earlier were not described for the given concrete type of the materials have been revealed. Among them one can call ordered bicontinuous double diamond, perforated lamellar layer, double gyroid structures, etc. The case of PCC is also attractive by the following reason. It is possible to embed metal ions in the complexes. After subsequent reduction of the ions metal nanoparticles with the defined size distribution will form in these materials [29-32]. The fact opens new possibilities regulating the catalytic activity of the metals.

It could be expected that the self-organizing processes are typical not only for biopolymers but for synthetic macromolecules consisting of pieces of the necessary form. The compounds having rigid tapered fragments in a lateral chain and simulating self-assembly processes of the tobacco mosaic virus have been synthesized on the basis of derivatives from gallic acid at the laboratory headed by Prof. V. Percec [33-35]. Since the early 1990s, the researches of the materials are carried out [36, 37]. It has been revealed that some compounds self-assemble in supramolecular cylinders which, in a block state, are organized in either a 2D ordered liquid crystal (LC) column phase or disordered one. Owing to some changes the chemical structure, for example, by increasing the number of the aliphatic tails or by their partial fluorinating, 3D mesophases are formed [16]. The transition from the disordered 2D hexagonal to 3D cubic mesophase of the $Ia\bar{3}d$ space group occurs in partially fluorinated samples of a macromonomer [38].

There are two basic approaches to the analysis of LC states.

The first one is usually utilized by the experts in area of WAXS and SAXS methods. They try to explain 1) ratios of intensities of reflections in thermo - and lyotropic LC, 2) the form of SAXS maxima. For example, when one studies a hexagonal phase, the structure of the LC state can suppose [39] to be presented as cylinders of a radius $R_1$ rounded by a hydrophobic shell from aliphatic radicals (for the case of surfactant/water) with a thickness $h = R_2-R_1$. Then the intensity of X - ray scattering, $I(s)$, proportional to the second power of the structure factor, $F(s)$, is expressed by the formula received by Oster and Riley in 1952 [39]

$$I(s) \sim F^2(s) \sim (\Delta\rho)^2 \, [(R_1 s J_1(R_1 s) - h s J_1(h s))/((R_1 s)^2 - (h s)^2)]^2.$$

Here, $\Delta\rho$ is the difference of densities between the cylinder and the shell,

$$s = 4\pi sin\theta/\lambda \qquad (2)$$

and *2θ* are the wave vector and the scattering angle, respectively [40]; *λ* is the wavelength.

Applicable only for the case of a well-distinguishable domain structure, concrete realizations of the similar concepts are used by various authors [8, 11, 15, 16].

The second approach in many respects complementary to the first is based on the fluctuation theory. By minimizing the Landau free energy functional [41, 42]

$$\Delta\Omega/T = \int(g|\nabla\psi|^2 + a_2\psi^2 + a_3\psi^3 + a_4\psi^4)dV \qquad (3)$$

on amplitudes and angles of a density fluctuation wave (*ψ ~ Δρ exp (ısr)*) [43, 44] one can find conditions of the stability for phases of either symmetry, points at which the transition from one phase to another can be observed, etc. Among the basic results received by the given method it is possible to name the following if $a_4$ does not depend on the angles between the wave vectors of the density fluctuations: 1) even the equality to zero of the parameter $a_3$ at the term of the third degree in the polynomial expansion of the free energy in 3D space results in that the phase transition an isotropic liquid – crystal is the 1-st order transition although it should be the 2-nd order transition in accordance with the Landau theory; 2) this transition can occur with the very low hidden heat as the cascade of the transitions I - BCC - HEX - LAM where BCC is a body centered cubic phase; 3) the direct transition I - LAM is possible; 4) other phases, such as orthorhombic, icosahedral, are non stable. It should be noted that the cascade of the transitions I - BCC - FCC was obtained theoretically by Kirgnits and Nepomnyashchii in 1970 (compare with (1) and see article [44] and references in it), and the theory of microphase separation in heteropolymers was developed by authors of the papers [45-47].

Now the approach involved is advanced for many cases [48, 49], though it is restricted to the requirement of a small value for the amplitude of the density fluctuation wave in order to the expansion (3) could be correct. Other theoretical methods are often unfit for the description of various polymer systems owing to the large relaxation times (see article [48] and references in it).

The aim of the present article is to analyze the PCC structure by help of experimental, theoretical and computer simulation methods and to discuss a possibility of the formation of bicontinuous structures in cubic mesophases. As shown below, the bicontinuous structures take place in the vicinity of the Lifshitz point. At this point the following structures can be coexist [42]: spatially modulated ordered, uniformly ordered and disordered. The reciprocal intensity becomes a linear function of the forth power of the scattering wave vector. The researches of LC, magnet, polyelectrolyte, block copolymer systems [50-53] have testified to the large role of fluctuations near the Lifshitz point. The recent studies have confirmed the conclusion. Critical exponents have been estimated by the analytical method of the $\varepsilon$–expansion [54] and the numerical Monte Carlo technique [55]. The exponents have differed from the mean-field values. Note that although the authors of the paper [53] have obtained the mean-field exponents, however, the phase diagram is different from the one predicted by the mean-field theory, so apparently some fluctuation corrections are necessary.

**EXPERIMENTAL AND CALCULATION METHODS**

**1. 3D MESOPHASES**

Owing to the large viscosity (in some thousands of times exceeding the viscosity of water) and the isotropy of optical (electrical and magnetic) properties 3D mesophases are named as viscous isotropic. Probably, in order to underline their 3D periodicity the mesophases are also called plastic. They have a cubic symmetry confirmed by optical polarizing microscopy and SAXS data. SAXS curves possess a system of narrow diffraction peaks [8, 12, 13, 16]. Sometimes [12, 13, 16], the positions of the most intensive maxima are in the ratio of square roots from 4, 5, 6 (Figure 1). The microscopic researches show no birefrequency. Therefore, the systems have a cubic symmetry.

Due to the same reasons the PCC structure of the cross-linked polymethacrylic acid (PMA) with cetyl pyridinium chloride (CPC) [29] should be referred to CP.

**2. CONSTRUCTION OF A MODEL AND DEFINITION OF SPACE GROUP**

As seen from SAXS curves [12, 13, 16, 29] or from the values of structure factors [8] sometimes it is very difficult to distinguish the first orders of reflections such as *(100), (110), (111)*. It means that the unit cell has a basis. BCC has an advantage in a choice of the initial assumptions before FCC and diamond lattices since in the last cases, respectively, the two maxima *(100), (110)* or one *(111)* have to remain from the first six reflections whereas in the first case only three *(100), (111)* and *(210)* must absent [40]. However, for BCC the peak *(110)* has significant intensity and there is no *(210)* reflection, the strongest one from presented in Figure 1. Some possibilities remain to define the positions of missing sites of the lattice and to result in an agreement with experimental data. However, the further considerations require data about geometric features of the structure of the complex.

Let us enter a vector directed along the hydrophobic surfactant radical to the polar fragment of the complex and take a small cube whose edge has a size much smaller than the parameter of the unit cell, and we shall fill the cube as it is possible more densely by repeated gel units together with surfactant counterions. It is clear that owing to a large difference in the energy of the interaction between polar and non polar parts of the complex a state, at which the hydrophobic surfactant radicals will situate parallel to four opposite in pairs faces of the small cube and perpendicular to the fifth one whereas the sixth face will become hydrophilic, can appear. Thus, we have introduced the vector field not only for one ionic pair (IP) but for the several IPs, occupying a certain volume (Figure 2a). It should be understood that the real IPs are formed in a poor polar medium [49]. Here and below we use the term to distinguish the well-organized regular structure of the complexes and the disordered one of swelling charged networks and weak surfactant solutions.

The vector field has the lowest symmetry, two mutually perpendicular rotary axes of the 1-st order belonging to the polar, for example, plane, therefore, it can not be presented in systems of a higher symmetry. Consequently, it is necessary to assume that the non polar face of the small cube should be pasted to the non polar face of another similar cube. In the given system there are already two mutually perpendicular rotary axes of the 2-nd order (Figure 2b). The rectangular prism with a square base, which has been received as a result of pasting these two cubes together, can go out on the face of the unit cell by one lateral surface (Figure 2c). The number of the prisms will be equal to 4 per one face, and some lattice sites of the $Pm\bar{3}n$ space group, which are in the special positions c or d [56], are on the prisms[1]. One can see

---

[1] As usually, the $Pm\bar{3}n$ space group is chosen instead of less symmetric but also possible the $P\bar{4}3n$ one.

that the symmetry of the sites does not break under the present construction (Figure 2).

## 3. TECHNIQUE OF CALCULATIONS

The quantitative evaluations of the number of structural elements or IPs and geometric characteristics of the PCC are received for some molar composition (PMA/CPC - 1:1 [29]) with a molecular weight $\mu = 564\ g\ mol^{-1}$, corresponding to the 50 wt.% water content, and with an average density $\rho = 1.2\ g\ cm^{-3}$ taking into account the proposed scheme (Figure 2). The density value is approximate and it is chosen from the experimental fact that the sample sinks in water. The real magnitudes, apparently, will not significantly differ from calculated, and all the corrections are introduced elementary. We have neglected by errors connected with a non homogeneous distribution of water over a gel body [14] and also with a possible non stoichiometric structure of the similar complexes [27, 28]. All these assumptions are used for the PCC of the cross-linked polydiallyldimethylammonium chloride with sodium decylsulfate (PDADMACL/DS) [13].

The number of structural elements or IPs, $N$, is equal

$$N = v\rho N_A/\mu \qquad (4)$$

where $v = a^3$ is the volume of the unit cell, which is a cube with the edge $a$, $\rho$ is the sample density, $N_A$ is the Avogadro number, $\mu$ is the molecular weight of the structural element. The aggregation number of the "spherical" micelles, $N_s$, which situate at the corners and the center of the unit cell or in the special positions a [56], makes approximately

$$N_s = (\rho N_A/\mu)\ (4\pi l^3/3) = \pi N/48 \qquad (5)$$

if to assume that their radius is equal to $l \approx 0.25a$ (Figures 2, 3). Then

$$N_c = (l^3 \rho N_A/\mu) = N/64 \qquad (6)$$

is the number of IPs belonging to some small cube with the edge of $l$, and, as its volume is equal to $l^3$, therefore, the interface surface area per one hydrophobic radical, $S$, is

$$S = l^2/N_c. \qquad (7)$$

Our assumptions are the following. The length of the aliphatic radical together with the thickness of the polymer gel branch and the polar fragment of the complex is equal to $l≈0.25a$ (Table). Some differences between these two parameters in the studied complexes and similar parameters received from the lamellar phase of aqueous solutions of cetyl trimethylammonium bromide and sodium dodecylsulfate [39] are reasonable. The total number of the small cubes, belonging to one face, is *8* and, consequently, the total number of the structural elements of the given type in the unit cell, $N_{ns}$, is

$$N_{ns} = 8 \times 6 \times N_c = 3N/4 \qquad (8)$$

where *6* is the number of the faces of the unit cell and we have used (6). The structures are named the "non spherical" micelles (Table). Subtracting $N_{ns}$ from $N$ and then from the obtained difference $2N_s$ (*2* is the number of the "spherical" micelles per the unit cell) we find by help of (5) and (8) that some number of IPs, $N_{bs}$,

$$N_{bs} = N - 2N_s - N_{ns} = (6 - \pi)/24 \qquad (9)$$

belongs, as shown below, to a bicontinuous structure.

All the original and calculated data are presented in Table.

In order to obtain the simulated SAXS curve (Figure 1b) we have used the expression for a convolution

$$I(s) \sim F^2 \, Z(s) * |\Sigma(s)|^2$$

of the functions $\Sigma(s)$, the Fourier transformant of the form factor (the lattice has been supposed cubic with the number of sites 1000), and $Z(s)$, the function taking into account the 2-nd order distortions [40], $F$ is the structure factor.

The maps of an electron density distribution (Figure 3) are calculated from experimental curves (Figure 1a, [13]) corrected by the Lorentz and multiplicity factors under the formula

$$\Delta\rho \sim \sum_{hkl} (\pm)|F_{hkl}| \, cos(2\pi hx)cos(2\pi ky)cos(2\pi lz)$$

which is applicable for the centrosymmetric lattice [16, 40], $|F_{hkl}|$ is the modulus of the structure factor. A combination of phases is selected "-+--" for the reflections *(200), (210), (120)* and *(211)*, respectively, taking into consideration the initial hypothesis (Figure 2) and results of the work [16]. Here "+" corresponds to the phase *(hkl)* equal to *0* and "-" stands for the phase equal to $\pi$.

To check the used methods the results of the work [13] are taken since some additional reflections are observed in the PCC of PDADMACL/DS. The phase set is chosen the following "-+++-+-+++--+", respectively, for the reflections *(110), (220), (310), (130), (222), (320), (230), (321), (231), (410), (140), (421), (241)*. The phase

set does not significantly change the electron density distribution maps obtained by help of the mentioned above phase combination. The first set together with the second one is applied to reconstruct the electron density from SAXS data.

**RESULTS AND DISCUSSIONS**

From the curve presented in Figure 1a one can see that after the desmeared correction on a height of the receiving slit a broad maximum [29] is divided into three narrow peaks whose positions are in the ratio of square roots from 4, 5, and 6. This fact allows to consider the unit cell as cubic with an edge *a = 10.4 nm* [40], the wave vector is determined under the formula (2), the wavelength is 0.1542 nm. The curve is similar to the represented in Figure 1b simulated curve especially in the field of wave vectors where the principal maxima are located. All differences between the calculated and experimental curves can be referred to the suggested model (Figure 2). The following assumptions have been made: there are 2 sites with a decreased electronic density in the center and one corner of the unit cell, i. e. in the special positions a, and there are 6 ones in the special positions c or d [56], the 2-nd order distortions are introduced to neglect the scattering at higher values of the wave vector.

Contour maps of an electronic density distribution (Figure 3), calculated from the experimental curves (Figure 1a, [13]), show that the initial model (Figure 2) is a rough approximation: owing to a mutual influence the "non spherical" micelles, which are situated in bisectors of perpendicular faces of the unit cell, i. e. in the special positions c or d [56], are deformed, their form differs from the one of prism and they are even divided in half by a layer of intermediate density. The "spherical" micelles are also deformed. Some discussion on this question can be found elsewhere [16]. Nevertheless, using the model (Figure 2) we have produced evaluations for the content of IPs in the structures mentioned above and for sizes of the various structural elements in the complexes.

Paying attention on the fact that Figure 3, *z=0.25* indicates the existence of regions with the increased electronic density close to points occupying the special positions e [56] and lying on a half distance from the corners of the unit cell up to its center it is necessary to assume that the IPs belong to the bicontinuous structure, i. e. to saddle surfaces of the zero average curvature and negative Gauss one (see below). The result is unusual since in this case the hydrophobic surfactant tails can contact with the polar shell of the "spherical" micelles. As a result, the free energy of the structures must increase. The expression (11) shows that it is possible.

Let us consider the case of the PMA/CPC complex as an example. The estimations for the PDADMACL/DS complex are calculated analogously. For the bicontinuous structure one direction can have *3.6 ≅ 172/8/6* of IPs (Table) where *8* is the number of the corners and *6* is the amount of the various directions equal to double space dimension. In this way dividing the $N_{ns}$ by *6*, we define that *180.1 ≅ 1080.6/6* of IPs belong to one "non spherical" micelle. Taking into account that *180.1 > $N_s$* it is possible to suppose that the "non spherical" micelles are energetically more favorable than "spherical". Thus we can explain the sign "-" in the formula (11).

By analyzing the results (Figure 3) one can write the free energy sensing the structure, $\omega$, as following:

$$\omega = 2f + 6f_1 + 8f_2, \qquad (10)$$

where *2*, *6* and *8*, respectively, are the number of both the "spherical" and "non spherical" micelles and the "knots" of the bicontinuous structure per the unit cell, and $f$, $f_1$ and $f_2$ are the free energies of the structures. The authors of the work [44] have really shown that the direct 1-st order transition I - BCC is possible. The phase called by them as $BCC_2$ is absolutely stable depending on relations between the 4-th degree structural correlators, if the latter are functions of the angles between the wave vectors of the density fluctuations. The energy of $BCC_2$ is counted from the level of LAM and the parameter $a_3$ in the functional (3) can be equaled to zero. Let us write (10) in designations of the work [44]:

$$\omega(BCC_2) = (\lambda - 3\lambda_1 + 4\lambda_2) A^2, \ a_n^2 = A/6, \qquad (11)$$

where $\lambda$, $\lambda_1$ and $\lambda_2$ are the structural correlators, $\lambda$ does not depend on the angles between the wave vectors, $\lambda_1$ and $\lambda_2$ depend on only one and two angles, respectively, $a_n$ is the amplitude of the density fluctuation wave. Comparing (10) and (11) we shall receive the relations i) $f = \lambda A^2/2$ for the "spherical" micelles where it is possible to regard $\lambda = const$, ii) $f_1 = - \lambda_1 A^2/2$ for the "non spherical" micelles (the result requires the explanation though the experimental data support it, as the greatest number of IPs in the unit cell belongs to the micelles of the given type) and iii) $f_2 = \lambda_2 A^2/2$ for the bicontinuous structure. Substituting the expression (5), (8) and (9), respectively, for $f$, $6f_1$ and $8f_2$ into (10) and taking into account (11) we find $\omega/N$:

$$\omega/N = (\pi N/24 - 3N/4 + (6 - \pi)N/24)/N = - 0.5. \qquad (12)$$

However, to pass from the functional (3) to the analysis (10), (11), it is necessary to make one more supposition: let us add to (3) the term

$$\int n(\mathbf{r}) (M^2/(2\chi) - MH)dV,$$

which after a variation with respect to $M$ transforms to the integral

$$\int n(\mathbf{r}) (-MH/2)dV. \tag{13}$$

Here $M$ and $H$ are the conjugate fields, $M = \chi H$, $H$ is the effective external field, $\chi$ is the "susceptibility", $n(\mathbf{r})$ is the local concentration of IPs, expressed in their number per the total amount in the unit cell. Supposing $H = \beta l \nabla \psi$ ($\beta$ is a dimensional coefficient, $l$ is the length scaled the value of space heterogeneity) it is possible to notice from (3) and (13) that at some concentration $n^*$ the terms (13) and $g|\nabla\psi|^2$ can mutually compensate each other. If it will happen near $a_2 = 0$ then the point is named as the Lifshitz one [42]. In it the 1-st order phase transition line is intersected by the 2-nd order transition line. On the other hand, the analysis of the authors of the work [44], carried out for the case in which the parameter $a_3$ is equal to zero, allows to write, if $n > n^*$:

$$\Delta\Omega/T = \int (-nMH/2 + a_2\psi^2 + a_4\psi^4)dV, \tag{14}$$

and at $a_2 < 0$, $H \to 0$ to receive the solution:

$$\psi = \pm(-a_2/2a_4)^{1/2} + \varphi(Z - X^2 + Y^2),$$

where the coordinates $X, Y, Z$ are counted from the "knots" of the bicontinuous structure, and about their origin we have the identity:

$$g\Delta\psi = g(\partial^2\varphi(0)/\partial Z^2 - 2\partial\varphi(0)/\partial X + 2\partial\varphi(0)/\partial Y) = g\partial^2\varphi(0)/\partial Z^2 = 0,$$

since $Z = 0$ is the inflection point ($\varphi(Z) = -\varphi(-Z)$), as seen from the symmetry of the functional (14), and the average curvature $C \sim -(\partial\varphi(0)/\partial X)^2 + (\partial\varphi(0)/\partial Y)^2 = 0$ on saddle surfaces of the type of a hyperbolic paraboloid if $Z$ axis is directed along the normal to the surface $\varphi(Z - X^2 + Y^2) = 0$ ($\partial\varphi(0)/\partial X = \partial\varphi(0)/\partial Y$). The result is obtained by help of the differential geometry [57], the infinite periodic minimal surfaces are defined by the equation $\varphi(Z - X^2 + Y^2) = 0$ and have the double Schwarz minimal surface genus. The Schwarz minimal surface genus is equal to *3* [58].

Due to the equality to zero $a_3$ and the compensation of the terms $g|\nabla\psi|^2$ and (13) at $n = n^*$ the functional (3) may be performed as the expression (11) which has been obtained by the authors of the work [44]. They have drawn a conclusion that the direct 1-st order transition I - BCC$_2$ is possible when the 4-th degree structural correlators depend on the angles between the wave vectors of the density fluctuations.

Assuming $g \sim S$ we find from (3), (7) and (13)

$$\chi\beta^2 \sim S/(n^* l^2) \qquad (15)$$

where $n^*$ is the fraction of IPs belonging to the bicontinuous structure.

Since the "spherical" micelles and the bicontinuous structure have the positive free energy this allows to assume the following scenario. These structures are formed at the Lifshitz point through the 2-nd order phase transition, i. e. without breaking the symmetry of the system while the final structure is built at the second stage by the 1-st order phase transition. In the last case an effective field results in breaking the symmetry, the total free energy becomes negative due to the contribution of the "non spherical" micelles, the cubic symmetry appears in the system. Thus, the analytical fluctuation theory [43, 44] is confirmed. The "susceptibility" value or the reciprocal number of the structural elements of the bicontinuous structure is a measure of the distance to the Lifshitz point (Table).

## CONCLUSIONS

The simple hypothesis (Figure 2) permits not only to design the model applicable for the simulation of the scattering curve (Figure 1b) but allows to interpret the experimental results from a point of view of the bicontinuous structure formation: for sections by planes *z = 0.25, 0.75* in the regions close to the points *{0.25, 0.25}, {0.25, 0.75}, {0.75, 0.25}, {0.75, 0.75}* the electronic density is increased (Figure 3), i. e. there are many IPs at the points. The geometric characteristics of the complexes (Table) can be easily estimated by help of the present model (Figure 2). The combination of the theoretical and experimental methods together with computer simulations is the base which enables to achieve the results.

## ACKNOWLEDGEMENTS

The author is thankful to the Russian Foundation of Basic Research (Grants No 01-03-32225) for financial support.

Thanks are expressed to Prof. S.N. Chvalun from Moscow Institute of Physics and Technology for helpful discussions and to Dr. A.V. Mironov for the technical assistance in preparing the manuscript.

# References

1. B.A. Parviz, D. Ryan, G.M. Whitesides, *IEEE Trans. Adv. Pack.*, Vol. 26, 2003, pp. 233-241.
2. J.M. Lehn, *Science*, Vol. 227, 1985, pp. 849-856.
3. J.M. Lehn, *Angew. Chem. Int. Ed. Engl.*, Vol. 27, 1988, pp. 89-112.
4. H. Ringsdorf, B. Schlarb, J. Venzmer, *Angew. Chem.*, Vol. 100, 1988, pp. 117-162.
5. H.J. Schneider, H. Durr, *Frontiers in Supramolecular Organic Chemistry and Photochemistry*, VCH, New York, 1993.
6. M. Shibayama, T. Tanaka, *Adv. Polym. Sci.*, Vol. 109, 1993, pp. 1-62.
7. A. Klug, *Phil. Trans. Roy. Soc. London*, Vol. 348A, 1994, pp. 167-178.
8. P. Mariani, V. Luzzati, H. Delacroix, *J. Mol. Biol.*, Vol. 204, 1988, pp. 165-189.
9. R.R. Balmbra, J.S. Clunie, J.F. Goodman, *Nature*, Vol. 222, 1969, pp. 1159-1160.
10. L.E. Scriven, *Nature*, Vol. 263, 1976, pp. 123-125.
11. M. Antonietti, J. Conrad, *Angew. Chem. Int. Ed. Engl.*, Vol. 33, 1994, pp. 1869-1870.
12. H. Okuzaki, Y. Osada, *Macromolecules*, Vol. 28, 1995, pp. 380-382.
13. E.L. Sokolov, F. Yeh, A.R. Khokhlov, B. Chu, *Langmuir*, Vol. 12, 1996, pp. 6229-6234.
14. A.T. Dembo, A.N. Yakunin, V.S. Zaitsev, A.V. Mironov, S.G. Starodubtsev, A.R. Khokhlov, B. Chu, *J. Polym. Sci.; Part B; Polym.Phys.*, Vol. 34, 1996, pp. 2893-2898.
15. Y.K. Kwon, S.N. Chvalun, J. Blackwell, V. Percec, J.A. Heck, *Macromolecules*, Vol. 28, 1995, pp. 1552-1558.
16. V.S.K. Balagurusamy, G. Ungar, V. Percec, G. Johansson, *J. Am. Chem. Soc.*, Vol. 119, 1997, pp. 1539-1555.
17. E.L. Thomas, D.B. Alward, D.J. Kinning, D.C. Martin, D.L. Handlin, Jr., L.J. Fetters, *Macromolecules*, Vol. 19, 1986, pp. 2197-2202.
18. H. Hasegawa, H. Tanaka, K. Yomasaky, T. Hashimoto, *Macromolecules*, Vol. 20, 1987, pp. 1651-1662.

19. M.M. Disko, K.S. Liang, S.K. Behal, R.J. Roe, K.J. Jeon, *Macromolecules*, Vol. 26, 1993, pp. 2983-2986.
20. S. Förster, A.K. Khandpur, J. Zhao, F.S. Bates, I.W. Hamley, A.J. Ryan, W. Bras, *Macromolecules*, Vol. 27, 1994, pp. 6922-6935.
21. M.E. Vigild, K. Almdal, K. Mortensen, I.W. Hamley, J.P.A. Fairclough, A.J. Ryan, *Macromolecules*, Vol.31. 1998, pp. 5702-5716.
22. B.J. Dair, C.C. Honeker, D.B. Alward, A. Avgeropoulos, N. Hadjichristidis, L.J. Fetters, M. Capel, E.L. Thomas, *Macromolecules*, Vol.32, 1999, pp. 8145-8152.
23. Yu.V. Khandurina, A.T. Dembo, V.B. Rogacheva, A.B. Zezin, V.A. Kabanov, *Vysokomol. soed.*, Vol. 36A, 1994, pp. 235-240 (*Polym. Sci., Ser. A*, Vol. 36, 1994, 189).
24. M. Antonietti, J. Conrad, A. Thünemann, *Macromolecules*, Vol.27, 1994, pp. 6007-6011.
25. Yu.V. Khandurina, V.L. Alexeev, G.A. Evmenenko, A.T. Dembo, V.B. Rogacheva, A. B. Zezin, *J. Phys. II France*, Vol. 5, 1995, pp. 337-342.
26. B. Chu, F. Yeh, E.L. Sokolov, S.G. Starodoubtsev, A.R. Khokhlov, *Macromolecules*, Vol. 28, 1995, pp. 8447-8449.
27. A.V. Mironov, S.G. Starodoubtsev, A.R. Khokhlov, A.T. Dembo, A.N. Yakunin, *Macromolecules*, Vol. 31, 1998, pp. 7698-7705.
28. A.V. Mironov, S.G. Starodoubtsev, A.R. Khokhlov, A.T. Dembo, A.N. Yakunin, *Colloids Surf.*, Vol. 147A, 1999, pp. 213-220.
29. L.M. Bronstein, O.A. Platonova, A.N. Yakunin, I.M. Yanovskaya, P.M. Valetsky, A.T. Dembo, E.E. Makhaeva, A.V. Mironov, A.R. Khokhlov, *Langmuir*, Vol. 14, 1998, pp. 252-259.
30. D.I. Svergun, E.V. Shtykova, A.T. Dembo, L.M. Bronstein, O.A. Platonova, A.N. Yakunin, P.M. Valetsky, A.R. Khokhlov, *J. Chem. Phys.*, Vol. 109, 1998, pp. 11109-11116.
31. L.M. Bronstein, O.A. Platonova, A.N. Yakunin, I.M. Yanovskaya, P.M. Valetsky, A.T. Dembo, E.S. Obolonkova, E.E. Makhaeva, A.V. Mironov, A.R. Khokhlov, *Colloids Surf.*, Vol. 147A, 1999, pp. 221-231.
32. D.I. Svergun, E.V. Shtykova, M.B. Kozin, V.V. Volkov, A.T. Dembo, E.V. Shtykova, Jr., L.M. Bronstein, O.A. Platonova, A.N. Yakunin,

P.M. Valetsky, A.R. Khokhlov, *J. Phys. Chem. B*, Vol. 104, 2000, pp. 5242-5250.

33. V. Percec, J. Heck, D. Tomazos, F. Falkenberg, H. Blackwell, G. Ungar, *J. Chem. Soc. Perkin Trans. 1*, No. 22, 1993, pp. 2799-2811.
34. V. Percec, J. Heck, G. Johansson, D. Tomazos, *Makromol. Symp.*, Vol. 77, 1994, pp. 237-265.
35. V. Percec, G. Johansson, J. Heck, G. Ungar, S.V. Batty, *J. Chem.Soc. Perkin Trans. 1*, No. 13, 1993, pp. 1411-1420.
36. Y.K. Kwon, S. Chvalun, A.-I. Schneider., J. Blackwell, V. Percec, J.A. Heck, *Macromolecules*, Vol. 27, 1994, pp. 6129-6132.
37. S.N. Chvalun, J. Blackwell, J.D. Cho, Y.K. Kwon, V. Percec, J.A. Heck, *Polymer*, Vol. 39, 1998, pp. 4515-4522.
38. S.N. Chvalun, M.A. Shcherbina, I.V. Bykova, J. Blackwell, V. Percec, *Vysokomol. soed., Ser. A*, Vol. 44, 2002, pp. 2134-2143 (*Polym. Sci., Ser. A*, Vol. 44, 2002, 1281).
39. F. Husson, H. Mustacchi, V. Luzzati, *Acta Cryst.*, Vol. 13, 1960, pp. 668-677.
40. A.Guinier, *Théorie et technique de la radiocristallographie*, 2nd ed., Dunod, Paris, 1956.
41. L.D. Landau, *Zh. Eksp. Teor. Fiz.*, Vol. 7, 1937, pp. 627-632.
42. L.D. Landau, E.M. Lifshitz, *Statistical Physics, Part 1*, 4th revised ed., ed. by L.P. Pitaevskii, Nauka-Fizmatlit, Moscow, 1995.
43. S.A. Brazovskii, *Zh. Eksp. Teor. Fiz.*, Vol. 68, 1975, pp. 175-185 (*Sov. Phys. JETP*, Vol. 41, 1975, 85).
44. S.A. Brazovskii, I.E. Dzyaloshinskii, A.R. Muratov, *Zh. Eksp. Teor. Fiz.*, Vol. 93, 1987, pp. 1110-1124 (*Sov. Phys. JETP*, Vol. 66, 1987, 625).
45. L. Leibler, *Macromolecules*, Vol. 13, 1980, pp. 1602-1617.
46. I. Ya. Erukhimovich, *Vysokomol. soyed.*, Vol. 24A, 1982, pp. 1942-1949 (*Polym. Sci. USSR*, Vol. 24, 1982, 2223).
47. G.H. Fredrickson, E.J. Helfand, *J. Chem. Phys.*, Vol. 87, 1987, pp. 697-705.
48. I. Ya. Erukhimovich, A.R. Khokhlov, *Vysokomol. soed.*, Vol. 35A, 1993, pp. 1808-1818.
49. A.R. Khokhlov, E.E. Dormidontova, *Usp. Fiz. Nauk*, Vol. 167, 1997, pp. 113-128 (*Phys. Usp.*, Vol. 40, 1997, 109).
50. J.H. Chen, T.C. Lubensky, *Phys. Rev. A*, Vol. 14, 1976, pp.1202-1207.

51. Y. Shapira, C. Becerra, N.F. Oliveira, Jr., T. Chang, *Phys. Rev. B*, Vol. 24, 1981, pp. 2780-2806.
52. J.F. Joanny, L. Leibler, *J. Phys. France*, Vol. 51, 1990, pp. 545-557.
53. F.S. Bates, W. Maurer, T.P. Lodge, M.F. Schulz, M.W. Matsen, K. Almdal, K. Mortensen, *Phys. Rev. Lett.*, Vol. 75, 1995, pp. 4429-4432.
54. H.W. Diehl, M. Shpot, *Phys. Rev. B*, Vol. 62, 2000, pp. 12338-12349.
55. M. Pleimling, M. Henkel, *Phys. Rev. Lett.*, Vol. 87, 2001, pp. 125702 (1-4).
56. *International Tables for X-ray Crystallography*, Vol. A, Kluwer Publishers, Dordrecht, the Netherlands, 1995.
57. D. Hilbert, S. Cohn-Vossen, *Geometry and the Imagination*, Chelsea Publishers, New York, 1952.
58. M. Wohlgemuth, N. Yufa, J. Hoffman, E.L. Thomas, *Macromolecules*, Vol. 34, 2001, pp. 6083-6089.

Table. The estimations of structure characteristics of the PMA/CPC and PDADMACL/DS complexes (composition - 1:1, water content 50%wt., the rest explanations are in the paper text).

| Parameters | PMA/CPC | PDADMACL/DS |
|---|---|---|
| $\mu$, the molecular weight of a structural element, $g/mol$ | 564 | 544.5 |
| $a$, the parameter of the unit cell, $nm$ | 10.4 | 7.79 |
| $N$, the number of the structural elements per the unit cell, (4) | 1440.8 | 627.2 |
| The scale of the space heterogeneity | $l=a/4$ | $l=a/4$ |
| $l$, nm | 2.6 | 1.95 |
| $N_s=\pi N/48$, the aggregation number of a "spherical" micelle, (5) | 94.3 | 41.1 |
| $N_c=N/64$, the number of IPs in the cube with the edge $l$, (6) | 22.5 | 9.8 |
| $N_{ns}=3N/4$, the number of IPs belonging to "non spherical" micelles, (8) | 1080.6 | 470.4 |
| $N_{bs}=N-2N_s-N_{ns}=(6-\pi)N/24$, the number of IPs belonging to a bicontinuous structure, (9) | 171.6 | 74.7 |
| $N_{bs}$ per one "knot" and one direction | 3.57 | 1.56 |
| $\omega/N$, the free energy of the cubic mesophase, (12) | -0.5 | -0.5 |
| $2f/N=2N_s/N$, the free energy of the "spherical" micelles | 0.131 | 0.131 |
| $6f_1/N= -N_{ns}/N$, the free energy of the "non spherical" micelles | -0.75 | -0.75 |
| $8f_2/N=n^*=N_{bs}/N$, the free energy of the bicontinuous structure | 0.119 | 0.119 |
| $S$, the interface surface area per one hydrophobic surfactant radical, $nm^2$, (7) | 0.3 | 0.387 |
| $\chi\beta^2 \sim S/(n^*l^2)=64/N_{bs}$, the "susceptibility" of the bicontinuous structure, (15) | 0.373 | 0.857 |

**Figure captions**

Figure 1a. The experimental SAXS curve typical for the complex PMA/CPC [29] after desmeared correction on a height of the receiving slit.

Figure 1b. The simulated SAXS curve for the $Pm\bar{3}n$ phase with the parameter of the unit cell, $a$, equal to *10.4 nm*.

Figure 2. The model structure of the PCC unit cell with the $Pm\bar{3}n$ space group: a) a small cube with the edge of *0.25a* where *a* is the parameter of the unit cell, b) a prism consisting of the two small cubes; c) a front view on the unit cell; d) a cross-section by the plane passing the inversion center; e) a cross-section by the plane parallel to two faces of the cell and passing at a distance *0.25a* from one of them (here, arrows represent schematically not only the direction from the hydrophobic radical to IP, as in Figure 2a, b, but the bicontinuous structure).

Figure 3a. Contour maps of an electron density distribution in the unit cell of the PMA/CPC complex for a phase combination "-+--" of the reflections *(200), (210), (120)* and *(211)*, respectively. Here and below, sections by planes *z=0, z=0.25* and a projection on the plane *(111)* are presented; the plane *(111)* passes the point *{0.25, 0.25, 0.25}*; the white, red, yellow colors point out the high density, the light-green and green colors show the middle density and the rest colors designate the low density.

Figure 3b. Contour maps of an electron density distribution in the unit cell of the PDADMACL/DS complex for a phase combination "-+--" of the reflections *(200), (210), (120)* and *(211)*, respectively. An additional phase set is used (see explanations in the text of the paper).

Figure 1a.

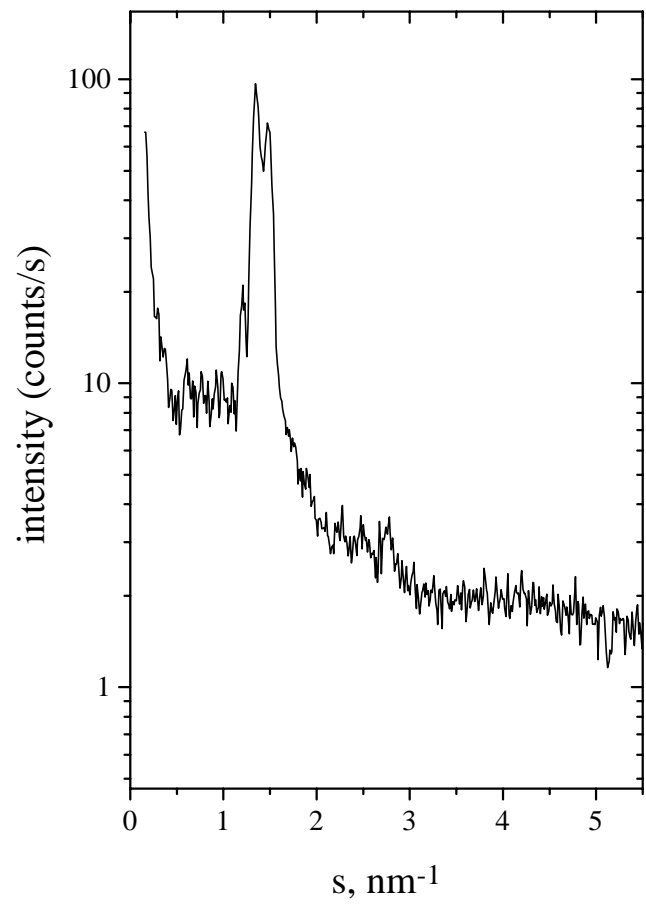

Figure 1b.

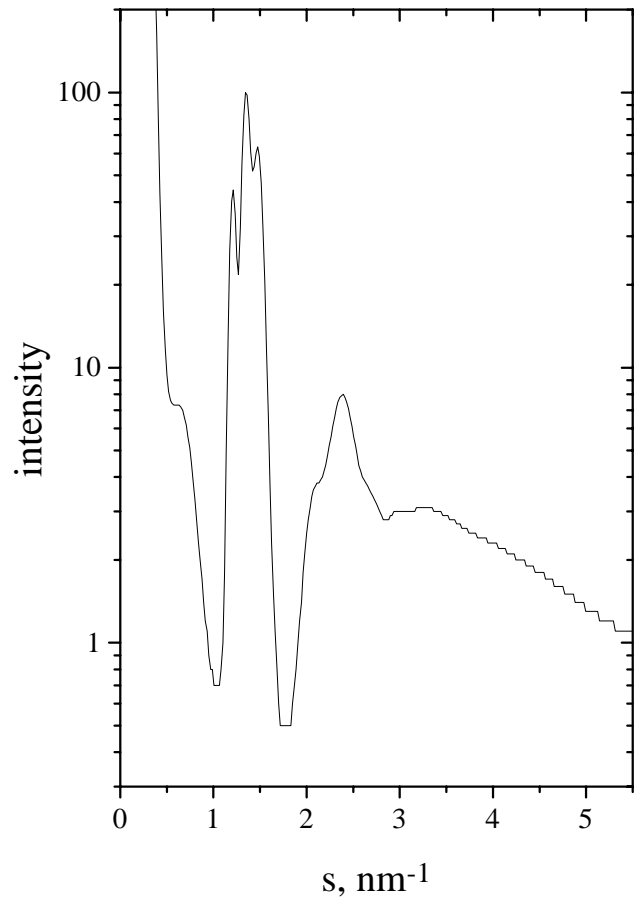

Figure 2.

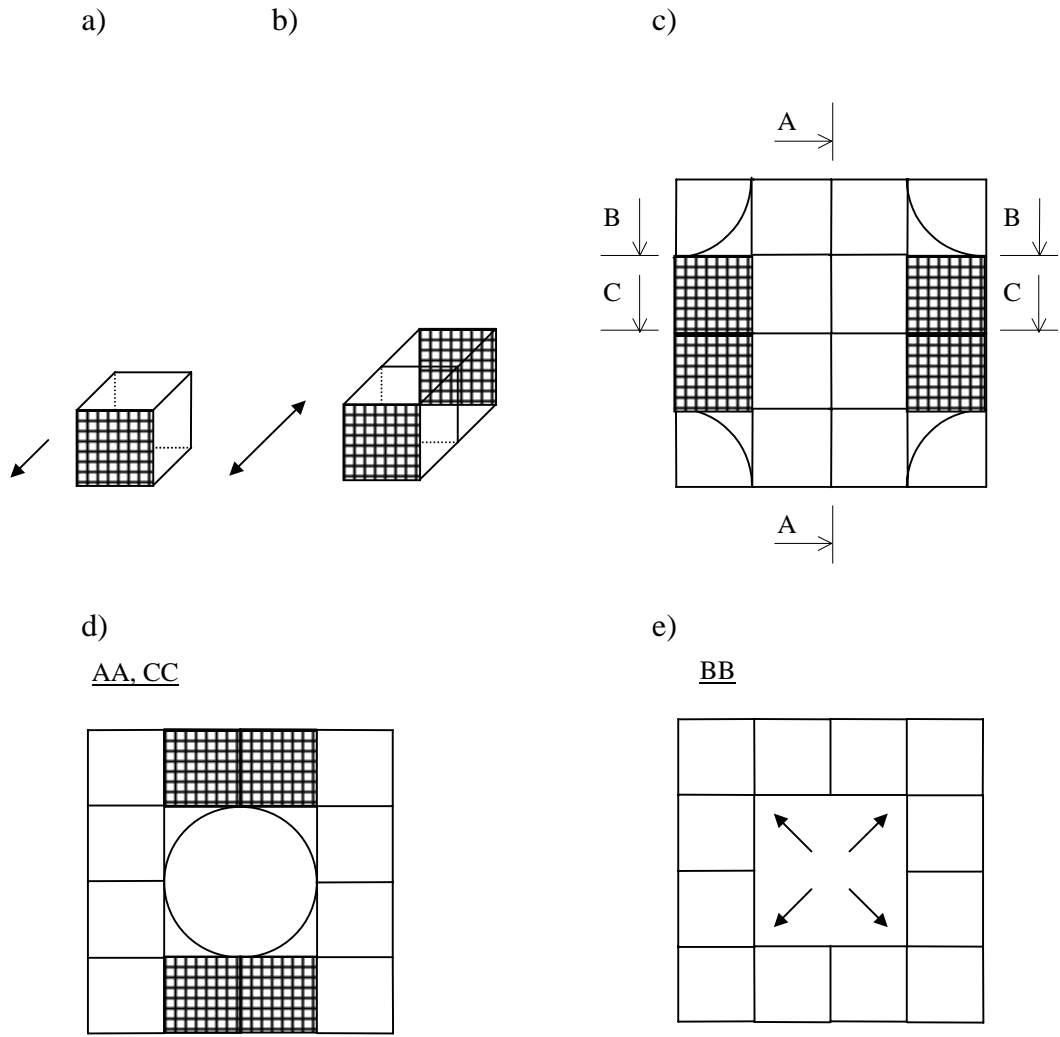

Figure 3a.

*z=0*

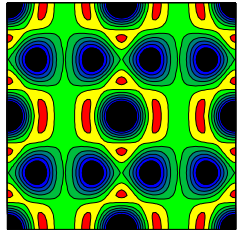

*z=0.25*

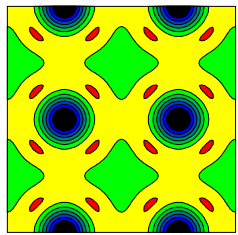

a projection on the plane *(111)*

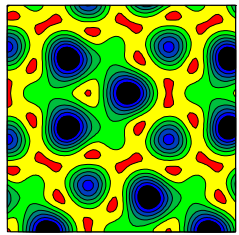

Figure 3b.

*z=0*

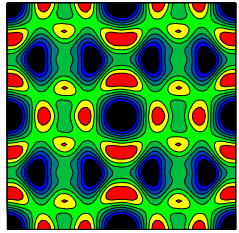

*z=0.25*

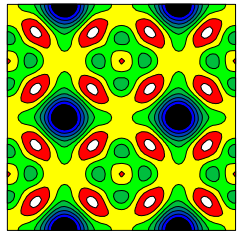

a projection on the plane *(111)*

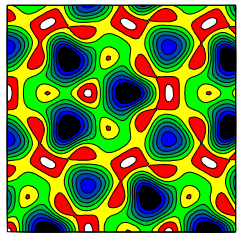